**Putting ChatGPT's Medical Advice to the (Turing) Test**

Authors: Oded Nov, PhD, MS[1,2], Nina Singh, BS[1], Devin M. Mann, MD, MS[1,3]

Affiliations: [1]NYU Grossman School of Medicine, Department of Population Health, New York, NY, USA; [2]Department of Technology Management, NYU Tandon School of Engineering, Brooklyn, NY, USA; [3]NYU Langone Health, Medical Center Information Technology, New York, NY, USA

Corresponding Author:

Oded Nov,

New York University

New York, NY, 11201

onov@nyu.edu

# Putting ChatGPT's Medical Advice to the (Turing) Test

**Abstract**

**Importance:** Chatbots could play a role in answering patient questions, but patients' ability to distinguish between provider and chatbot responses, and patients' trust in chatbots' functions are not well established.

**Objective:** To assess the feasibility of using ChatGPT or a similar AI-based chatbot for patient-provider communication.

**Design:** Survey in January 2023

**Setting:** Survey

**Participants:** A US representative sample of 430 study participants aged 18 and above was recruited on Prolific, a crowdsourcing platform for academic studies. 426 participants filled out the full survey. After removing participants who spent less than 3 minutes on the survey, 392 respondents remained. 53.2% of respondents analyzed were women; their average age was 47.1.

**Exposure(s):** Ten representative non-administrative patient-provider interactions were extracted from the EHR. Patients' questions were placed in ChatGPT with a request for the chatbot to respond using approximately the same word count as the human provider's response. In the survey, each patient's question was followed by a provider- or ChatGPT-generated response. Participants were informed that five responses were provider-generated and five were chatbot-generated. Participants were asked, and incentivized financially, to correctly identify the response source. Participants were also asked about their trust in chatbots' functions in patient-provider communication, using a Likert scale of 1-5.

**Main Outcome(s) and Measure(s):** Main outcome: Proportion of responses correctly classified as provider- vs chatbot-generated. Secondary outcomes: Average and standard deviation of responses to trust questions.

**Results:** The correct classification of responses ranged between 49.0% to 85.7% for different questions. On average, chatbot responses were correctly identified 65.5% of the time, and provider responses were correctly distinguished 65.1% of the time. On average, responses toward patients' trust in chatbots' functions were weakly positive (mean Likert score: 3.4), with lower trust as the health-related complexity of the task in questions increased.

**Conclusions and Relevance:** ChatGPT responses to patient questions were weakly distinguishable from provider responses. Laypeople appear to trust the use of chatbots to answer lower risk health questions. It is important to continue studying patient-chatbot interaction as chatbots move from administrative to more clinical roles in healthcare.

**Putting ChatGPT's Medical Advice to the (Turing) Test**

**Background**

Advances in large language models have enabled dramatic improvements in the quality of artificial intelligence (AI) generated conversations. Recently, the launch of ChatGPT[1] has prompted a surge in the public's interest in AI-based chatbots.[2,3] The present study assesses the feasibility of using ChatGPT or a similar AI-based chatbot for answering patient portal messages directed at healthcare providers. This application is of particular interest given the increasing burden of patient messages being delivered to providers[4] and the association between increased electronic health record (EHR) work and provider burnout.[5,6] Moreover, providers are generally not allocated time or reimbursement for answering patient messages.

In an age when patients increasingly expect providers to be virtually accessible, it is likely that patient message load will continue increasing. As the technology behind AI-based chatbots matures, the time is ripe for exploring chatbots' potential role in patient-provider communication.

Here, we report on the ability of members of the public to distinguish between AI- and provider-generated responses to patients' health questions. Further, we characterize participants' trust in chatbots' functions. Finally, we discuss the possible implications of adoption of AI-based chatbots in patient messaging portals.

**Methods**

Ten representative non-administrative patient-provider interactions from DM were extracted from the EHR. All identifying details were removed, and typos in the provider's response were fixed. Patients' questions were placed in ChatGPT with a request to respond, using approximately the same word count as the provider's response. Chatbot response text recommending consultation with the patient's healthcare provider were removed.

The ten questions and responses were presented to a US representative sample of 430 people aged 18 and above, recruited on Prolific, a crowdsourcing platform for academic studies.

Each patient's question was followed by either a provider- or ChatGPT-generated response. Participants were informed that five of the responses were written by a human provider and five by an AI-based chatbot. Participants were asked to determine which responses were written by the provider and which by chatbot. The order of the ten questions and answers, as well as the order of the choices presented to participants, were randomized. Participants were incentivized financially to distinguish between human and chatbot responses.

Participants were then asked questions about their trust in chatbots' use in patient-provider communication using a 1-5 Likert scale.

**Results**

426 participants filled out the full survey. After removing participants who spent less than 3 minutes on the survey, 392 survey responses were used in the analysis. 53.2% of the remaining respondents were women and their average age was 47.1 (16.0).

The responses to patients' questions varied widely in participants' ability to identify whether they were written by human or chatbot, ranging between 49.0% to 85.7% for different questions. Each participant received a score between 0-10 based on the number of responses they identified correctly (Figure 1). On average, chatbot responses were identified correctly in 65.5% of the cases, and human provider responses were identified correctly in 65.1% of the cases. No significant differences were found in response distinguishability or trust by demographic characteristics.

On average, patients trusted chatbots (Table 1), yet trust was lower as the health-related complexity of the task in question was higher. No significant correlations were found between trust in health chatbots and demographics or ability to correctly identify chatbot vs human responses.

**Discussion**

Patients increasingly expect "consumer grade" healthcare experiences that mirror their experiences with the rest of their digital life. They want omnichannel and interactive communication, frictionless access to care, and personalized education. The resulting overwhelming volume of patient portal messages highlights an opportunity for chatbots to assist healthcare providers. However, whether patients view chatbot communication as comparable to communication with human providers requires empirical investigation.[7-9]

In this study of a US representative sample, compared to the benchmark of 50% representing random distinguishability, and 100% representing perfect distinguishability, laypeople found responses from an AI-based chatbot to be weakly distinguishable from those from a human provider. Notably, there was very little difference between the distinguishability rate of human vs. chatbot response (65.5 vs. 65.1%). It is likely that in the near future, the level of indistinguishability we found will represent a lower bound of performance, as medically-trained chatbots will likely be less distinguishable. Another possible future development is for chatbots to reach superhuman level as seen in other medical domains.[10]

Respondents' trust in chatbots' functions were mildly positive. Notably, there was a lower level of trust in chatbots as the medical complexity of the task increased, with the highest acceptance being administrative tasks like scheduling appointments and the lowest acceptance being providing treatment advice. This is broadly consistent with prior studies.[11]

Identifying appropriate scenarios for deploying healthcare chatbots is an important next step. While chatbots in healthcare administrative tasks (e.g. scheduling) are widely used, optimal clinical use cases are still emerging.[12] Chatbots have been developed and deployed for highly specialized clinical scenarios such as symptom triage and post-chemotherapy education.[13] More generalized chatbots like ChatGPT represent a new opportunity to use chatbots in support of more common chronic disease management for conditions such as hypertension, diabetes and asthma. For example, chatbots could be deployed with home blood pressure monitoring to support patient questions about treatment plans, medication titrations and potential side effects.[14]

The findings suggest that in certain use cases, clinical chatbots will be acceptable. Potential models include chatbots that directly interact with patients (e.g., through patient portals) or serve as clinician assistants, generating draft text or transforming clinician documentation into more patient friendly versions. For providers' work, this would entail more *curation* and less *creation* of healthcare advice in response to virtual patient messages.

The appropriateness of each model might depend on the clinical complexity and severity of the condition. Higher risk/complexity clinical interactions would use chatbots to generate drafts for clinician editing/approval and lower risk situations may allow for direct patient-chatbot interaction. Alternatively, it may be useful to have chatbots classify questions into administrative versus health, replying directly to administrative ones and drafting responses for provider approval to health questions. The role and impact of disclosure of origination (human vs chatbot) also needs further exploration.

While our study addressed new questions with state-of-the-art technology, it has some key limitations. First, ChatGPT was not trained on medical data and could be inferior to medically-trained chatbots like Med-PaLM.[15] Second, there was no specialized prompting of ChatGPT (e.g. to be empathetic), which can help responses sound more human. Finally, this study used only ten real-world questions with human responses from one provider. Further studies incorporating larger numbers of real-world questions and responses are warranted.

In addition, future research may explore how to prompt chatbots to provide optimal patient experience, exploring if there are types of questions that chatbots are better at answering than others, and exploring if patients feel more trusting if there is clinician review before chatbots respond.

## Conclusion

Overall, our study shows that ChatGPT responses to patient questions are weakly distinguishable from provider responses. Furthermore, laypeople trusted chatbots to answer lower risk health questions. It is important to continue studying how patients interact (objectively and emotionally) with chatbots as they become a commodity and move from administrative to more clinical roles in healthcare.

## Acknowledgments

None.

## Funding

The authors receive financial support from the US National Science Foundation (awards no. 1928614 and 2129076) for the submitted work. The funding source had no further role in this study.

**Figures and Tables**

**Figure 1. Distribution of Correct Responses.**

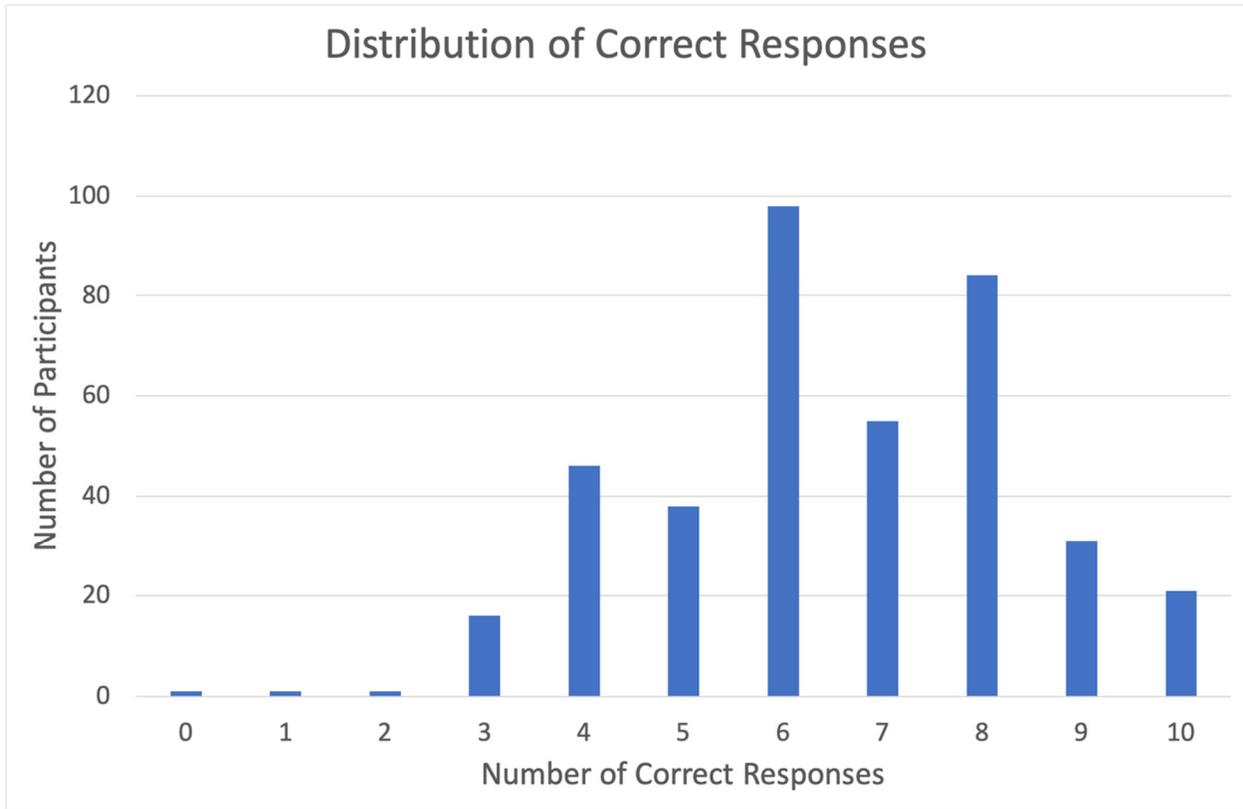

Each participant received a score between 0-10 based on the number of responses they identified correctly.

**Table 1. Health Chatbot Trust Questions and Responses.**

| Question | % of Patients with Likert Response of >= 4 | Mean Likert Response (1-5) and standard deviation |
|---|---|---|
| I could trust answers from a health chatbot about logistical questions (such as scheduling appointments, insurance questions, medication requests). | 79.6 | 3.94 (0.92) |
| I could trust a chatbot to provide advice about preventative care, such as vaccines, or cancer screenings | 63.3 | 3.52 (1.10) |
| I could trust a chatbot to provide diagnostic advice about symptoms. | 38.8 | 2.90 (1.14) |
| I could trust a chatbot to provide treatment advice. | 38.3 | 2.89 (1.12) |
| AI chatbots can be a more trustworthy alternative to Google to answer my health questions. | 59.2 | 3.56 (1.02) |
| Health chatbots could help me make better decisions. | 60.2 | 3.49 (0.91) |